\documentclass[prb,twocolumn,showpacs]{revtex4-1}

\usepackage{graphicx}
\usepackage{amsmath}
\usepackage{amssymb}
\usepackage{epsfig}
\usepackage{epstopdf}
\usepackage{gensymb}
\usepackage[sort&compress]{natbib}
\usepackage{wrapfig}

\begin{document}

\title{Fractional fractal quantum Hall effect in graphene superlattices}

\author{L. Wang$^{1,2}$}
\author{Y. Gao$^{1}$}
\author{B. Wen$^{3}$}
\author{Z. Han$^{3}$}
\author{T. Taniguchi$^{4}$}
\author{K. Watanabe$^{4}$}
\author{M, Koshino$^{5}$}
\author{J. Hone$^{1}$}
\author{C.R. Dean$^{3}$$^{\ast}$}

\affiliation{$^{1}$Department of Mechanical Engineering, Columbia University, New York, NY 10027, USA}
\affiliation{$^{2}$Department of Electrical Engineering, Columbia University, New York, NY 10027, USA}
\affiliation{$^{3}$Department of Physics, Columbia University, New York, NY 10027, USA}
\affiliation{$^{4}$National Institute for Materials Science, 1-1 Namiki, Tsukuba 305-0044, Japan}
\affiliation{$^{5}$Department of Physics, Tohoku University, Sendai, 980-8578, Japan}

%\begin{abstract}

%\end{abstract}

\maketitle

\noindent \textbf{  The Hofstadter energy spectrum provides a uniquely tunable system to study emergent topological order in the regime of strong interactions.  Previous experiments, however, have been limited to the trivial case of  low Bloch band filling where only the Landau level index plays a significant role. Here we report measurement of high mobility graphene superlattices where the complete unit cell of the Hofstadter spectrum is accessible. We observe coexistence of conventional fractional quantum Hall effect (QHE) states together with the integer QHE states associated with the fractal Hofstadter spectrum.  At large magnetic field, a new series of states appear at fractional Bloch filling index. These fractional Bloch band QHE states are not anticipated by existing theoretical pictures and point towards a new type of many-body state.}

In a 2D electron gas (2DEG) subjected to a magnetic field,  the Hall conductance is generically quantized whenever the Fermi energy lies in a gap\cite{TKNN:1982}.  The integer quantum Hall effect (IQHE) results from the cyclotron gap that separates the Landau energy levels (LLs). The longitudinal resistance drops to zero, and the Hall conductance develops plateaus quantized to $\sigma_{XY}=\nu e^{2}/h$, where $\nu$, the Landau level filling fraction, is integer valued. When the 2DEG is modified by a spatially-periodic potential, the LL's develop additional subbands separated by mini-gaps,  resulting in the fractal energy diagram known as the Hofstadter butterfly\cite{Hofstadter:1976}. When plotted against normalized magnetic flux, $\phi/\phi_{o}$,  and normalized density, $n/n_{o}$, representing the magnetic flux quanta and electron density per unit cell of the superlattice, respectively, the fractal mini-gaps follow linear trajectories\cite{Wannier:1978} according to a Diophantine equation, ${n}{/n_{o}}=t{\phi}/{\phi_{o}} + s$, where $s$ and $t$ are integer valued.  $s$ is the Bloch band filling index associated with the superlattice and $t$ is a similar index related to the gap structure along the field axis\cite{Macdonald:1983} (in the absence of a superlattice, $t$ reduces to the LL filling fraction). The fractal mini-gaps give rise to QHE features at partial Landau level filling, but in this case $t$, rather than the filling fraction determines the quantization value\cite{Streda:1982,TKNN:1982}, and the Hall plateaux remain integer valued.

In very high mobility 2DEGs, strong Coulomb interactions can give rise to many-body gapped-states also appearing at partial Landau fillings\cite{Tsui:1982,Laughlin:1983,Dean:2011}. Again the Hall conductance exhibits a plateau, but in this case quantized to fractional values of $e^2/h$. This effect is termed the fractional quantum Hall effect (FQHE), and represents an example of emergent behaviour in which electron interactions give rise to collective excitations with properties fundamentally distinct from the fractal IQHE states. A natural theoretical question arises regarding how  interactions manifest in a patterned 2DEG\cite{Gudmundsson:1995, Doh:1998,ChenUK:2014,Apalkov:2014}. In particular, since both the FQHE many-body gaps, and the single-particle fractal mini-gaps, can appear at the same filling fraction, it remains unclear whether the FQHE is even possible within the fractal Hofstadter spectrum\cite{Kol:1993,Pfannkuche:1997,Ghazaryan:2014}. Experimental effort to address this question has been limited owing to the requirement of imposing a well-ordered superlattice potential while preserving a high carrier mobility\cite{Schlosser:1996,Albrecht:2001,Melinte:2004}.

Here we report a low temperature magnetotransport study of fully encapsulated h-BN/graphene/h-BN heterostructures, fabricated by  van der Waals (vdW) assembly with edge contact\cite{Wang:2013} (see Methods). A key requirement to observe the Hofstadter butterfly is the capability to reach the commensurability condition in which the magnetic length, $l_{B}=\sqrt{\hbar/eB}$ ($\hbar$ is Planck's constant, $e$ the electron charge and $B$ the magnetic field) is comparable to the wavelength of the spatially periodic potential, $\lambda$. For experimentally accessible magnetic fields this requires a superlattice potential with wavelength of order tens of nanometers.  In this regard Graphene/h-BN heterostructures provide an ideal system since at near zero angle mismatch, the slight difference in lattice constants between the graphene and BN crystal structures gives rise to a moir\'{e} superlattice with a period of $\sim$14~nm\cite{Yankowitz:2012,Dean:2013,Ponomarenko:2013,Hunt:2013}. Moreover, we find that in our vDW assembly technique, in which the graphene/BN interface remains pristine \cite{Wang:2013}, alignment between the graphene and BN can be achieved by simple application of heat. Fig. 1c shows an example of a heterostructure that was assembled with random (and unknown) orientation of each material.  After heating the sample to $\sim350$~$^{o}$C, the graphene flake translates and rotates through several microns, despite being fully encapsulated between two BN sheets. This behaviour has been observed in several devices, in each case resulting in a moire wavelength ranging between $10-15$~nm (indicating less than $2^{o}$ angle mismatch\cite{Yankowitz:2012}). We postulate that the thermally induced motion proceeds until the macro-scale graphene flake finds a local energy minimum, corresponding to crystallographic alignment to one of the BN surfaces\cite{Woods:2014}, however, a full discussion of the mechanics of this process is beyond the scope of the present manuscript and will be the focus of a future study.
 
\begin{figure*}[t]
\begin{center}
\includegraphics[width=1\linewidth]{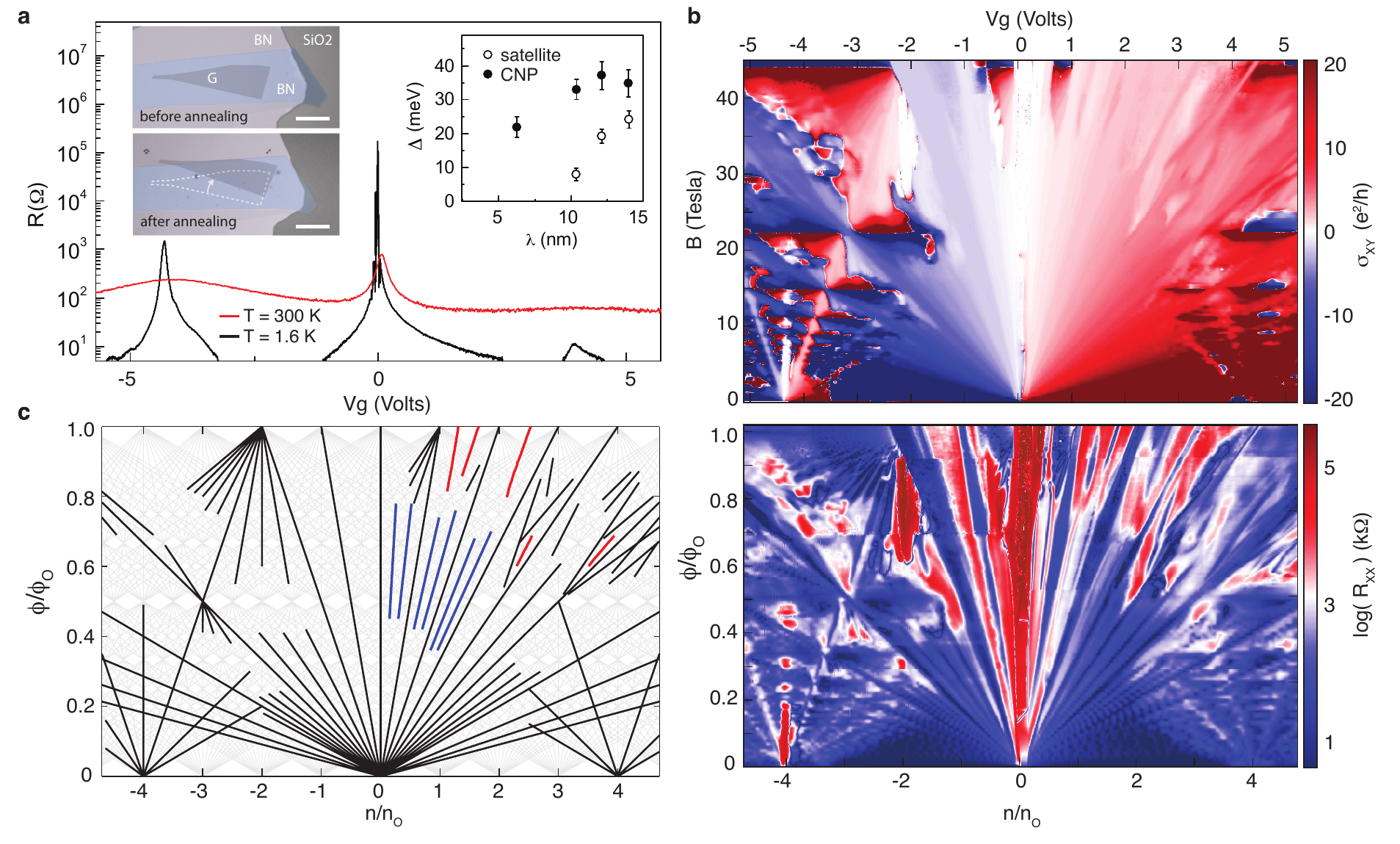}
\caption{(a) Zero field resistance versus gate bias measured in a BN encapsulated graphene device with a $\sim$10~nm moir\'{e} superlattice.  Inset left is a false-coloured optical image showing macroscopic motion of the graphene after heating.  Inset right shows the gap measured by thermal activation at the CNP and hole satellite peak positions, across 4 different devices (see SI). Both gaps are observed to vary continuously with rotation angle.  (b) Hall conductance plotted versus magnetic field and gate bias (upper panel) and longitudinal resistance versus normalized field and density (bottom panel) for the same device as in (a).  (c) Simplified Wannier diagram labelling the QHE states identified in (b). Families of states are identified by colour as follows:  Black lines indicate fractal IQHE states within the conventional Hofstadter spectrum, including complete lifting of spin and valley degrees of freedom.  Blue lines indicate conventional FQHE states. Red lines indicate anomalous QHE states exhibiting integer Hall quantization, but corresponding to a fractional Bloch index (see text)}
\label{fig:Fig1}
\end{center}
\end{figure*}

Fig. 1a shows the resistance versus density at zero applied magnetic field for a device with moir\'{e} wavelength $\sim10$~nm (see SI for more details). In addition to the usual peak in resistance at the zero-density charge neutrality point (CNP),  two additional satellite peaks appear at equidistant positive and negative gate bias $-$ characteristic signatures of electronic coupling to a moir\'{e} superlattice\cite{Yankowitz:2012,Dean:2013,Ponomarenko:2013}. The CNP peak resistance exhibits thermally activated behaviour, and exceeds 100~k$\Omega$ at low temperature, indicating a moir\'{e}-coupling induced bandgap\cite{Hunt:2013,Woods:2014,Amet:2015}. The gap varies continuously with rotation angle, consistent with previous studies of non-encapsulated graphene\cite{Hunt:2013}. At zero angle the energy gap measured by transport is equivalent to the optical gap\cite{ChenTLH:2014}, providing further indication of the high quality device realized by vdW assembly.  Unlike previous studies of encapsulated devices\cite{Ponomarenko:2013,Woods:2014},  we find that the gap remains robust despite the graphene being covered with a top BN layer. The precise origin of the gap in h-BN/graphene heterostructures remains uncertain\cite{Jung:2015}, and further experimental and theoretical studies will be required to resolve the differences in the gap magnitude and correlation with twist angle that have been reported so far.

%The discrepancy may result from different fabrication techniques and/or variation in sample disorder. Our devices exhibit exceptional mobility, with evidence of ballistic transport observed over length scales in excess of 10~$\mu$m at low temperature\cite{Wang:2013}, and with the activation gap values comparable to optical measurements\cite{ChenTLH:2014}. We note however that several questions remain regarding the origin of the gap in BN/graphene heterostructures\cite{Jung:2015}. A full understanding of the effect of encapsulation, and the rotation angle dependence, will require further experimental and theoretical study.

Fig. 1b shows the longitudinal resistance and transverse Hall conductance for the same device as in Fig. 1a.  The high quality of the device reveals a rich complexity in the transport signatures.  A sequence of repeated mini-fans, resembling a repeated butterfly in the hall conductance map, show clear evidence evidence of the fractal nesting expected from the Hofstadter spectrum. In Fig. 1c, a simplified Wannier diagram is shown in which the positions of the most prominent QHE states are plotted against normalized flux and normalized density axes. 
 Light grey lines indicate all possible gap trajectories according to the Diophantine equation, where we have assumed that both spin and valley degeneracy may be lifted such that $s$ and $t$ are allowed to take any integer value (for clarity the range is restricted to $|{s,t}|=0...10$). Black, blue and red lines trace QHE features identified in our experimental data according to the usual requirement of observing a minimum in longitudinal resistance concomitant with a quantized plateau in the Hall conductance.  Three distinct families of states are identified, corresponding to generalized IQHE states associated with the Hofstadter spectrum (black lines); conventional FQHE sates (blue lines), and anomalous sates that do not fit within either of these descriptions (red lines).  In the remainder of this manuscript we focus our discussion on the FQHE and anomalous sates.

 \begin{figure}[t]    
 \includegraphics[width=0.8\linewidth]{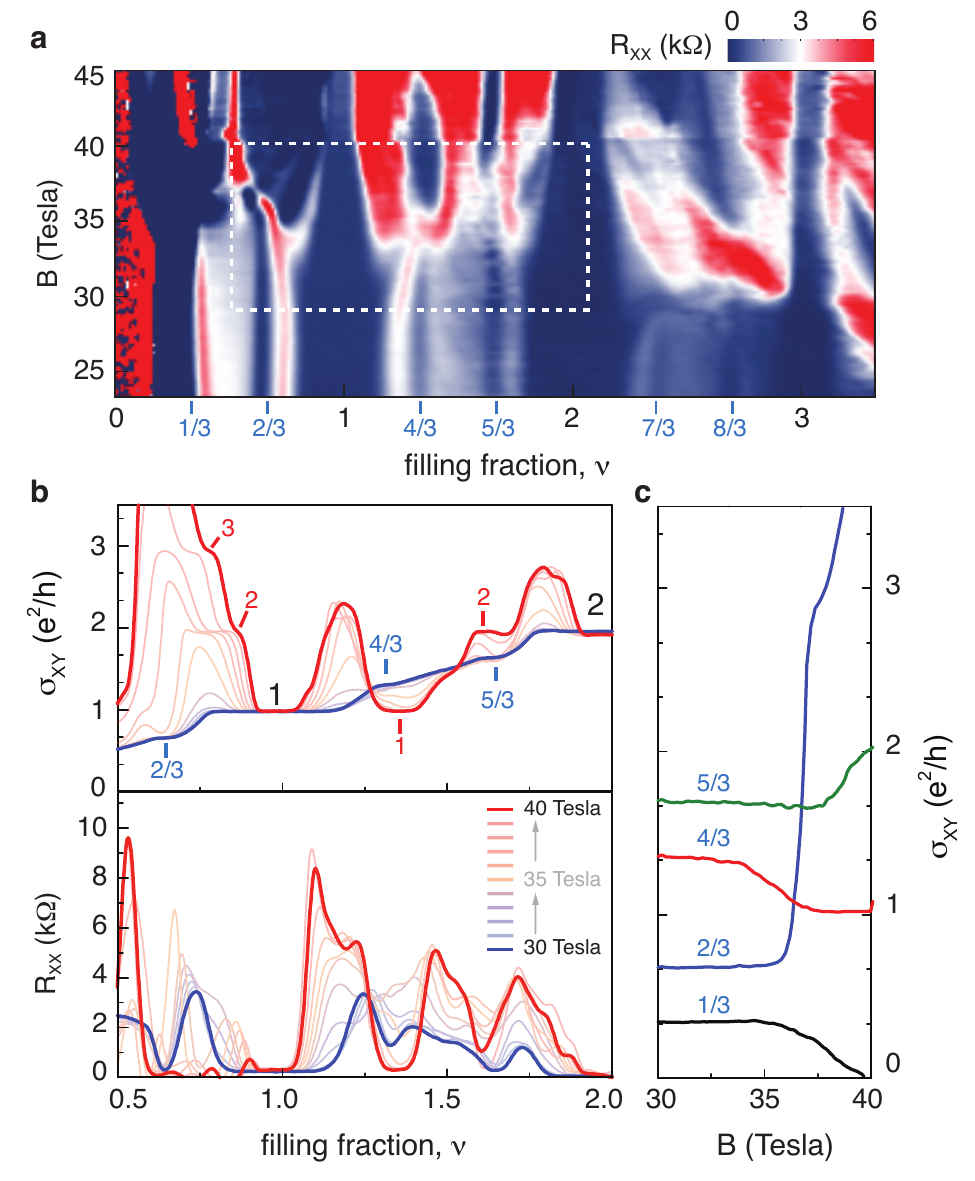}
		\caption{(a)Longitudinal resistance versus Landau filling fraction corresponding to a high field region of data from Fig. 1b.  (b) Hall conductance (upper panel) and longitudinal resistance (lower panel) corresponding to horizontal line-cuts within the dashed region in (a).  Conductance plateaus identified in the low field and high field limits are labelled blue and red, respectively (a).  Hall conductance versus magnetic field at fixed filling fraction showing evolution from FQHE plateaus to integer value plateaus.}
		\label{fig:Fig2}
\end{figure}

FQHE states are characterized by a longitudinal resistance minimum occurring at a fractional Landau filling index,  with the corresponding Hall conductance plateau quantized to the same fractional value, and with the gap trajectory in the Wannier diagram projecting to $n/n_{o}=0.$ The FQHE states are observed at all $m/3$ filling fractions in the lowest and first excited Landau level, where $m$ is integer valued.  The observation of a well developed 5/3 state is consistent with previous studies of monolayer graphene in which a zero field bandgap was reported\cite{Hunt:2013,Amet:2015}, and is presumably due to the lifting of the valley degeneracy that results from coupling to the moir\'{e} pattern\cite{ChenUK:2014}.

From Fig. 1c it can be seen that the FQHE states span only a finite range of perpendicular magnetic field.  In Fig. 2a, this is shown in more detail where a selected region of the longitudinal resistance from Fig. 1b, is replotted against magnetic field on the vertical axis and  Landau filling fraction along the horizontal axis. Select line traces from this diagram, corresponding to varying filling fraction at constant magnetic field, are additionally shown in Fig. 2b. For simplicity we focus on the 4/3 state as a representative example of the general behaviour. At $B=30$~T the Hall conductance at filling fraction 4/3 is well quantized to $\sigma_{XY}=4/3(e^{2}/h)$.  Upon increasing to $B=34$~T this state has completely disappeared, and by $B=40$~T a Hall plateau has reemerged, but quantized to a new value $\sigma_{XY}=1(e^{2}/h)$. In the high field state there is also a shift in the apparent filling fraction, with the integer plateau no longer coincident with $4/3$ filling (Fig. 2b).  We interpret the apparent phase transition to be the result of a competition with a fractal mini-gap state.  This is supported by examining the relative strength of the QHE features on either side of the transition as a qualitative measure of relative gap size; the high-field integer-valued state exhibits a significantly better-developed longitudinal resistance minimum, and wider Hall plateau (indicative of a larger gap) than the lower-field FQHE state.   Taken together, these observations provide experimental evidence supporting two theoretical predictions\cite{Kol:1993,Pfannkuche:1997,Ghazaryan:2014}:  (i) the fractal Hofstadter spectrum can support Laughlin-like FQHE states even at field strengths approaching the commensurability condition (ii) at filling fractions in which a conventional FQHE and a Hofstadter mini-gap state coexist, the state with the larger associated bandgap is the one which emerges.

\begin{figure*}[t]
	\begin{center}
	\includegraphics[width=1\linewidth]{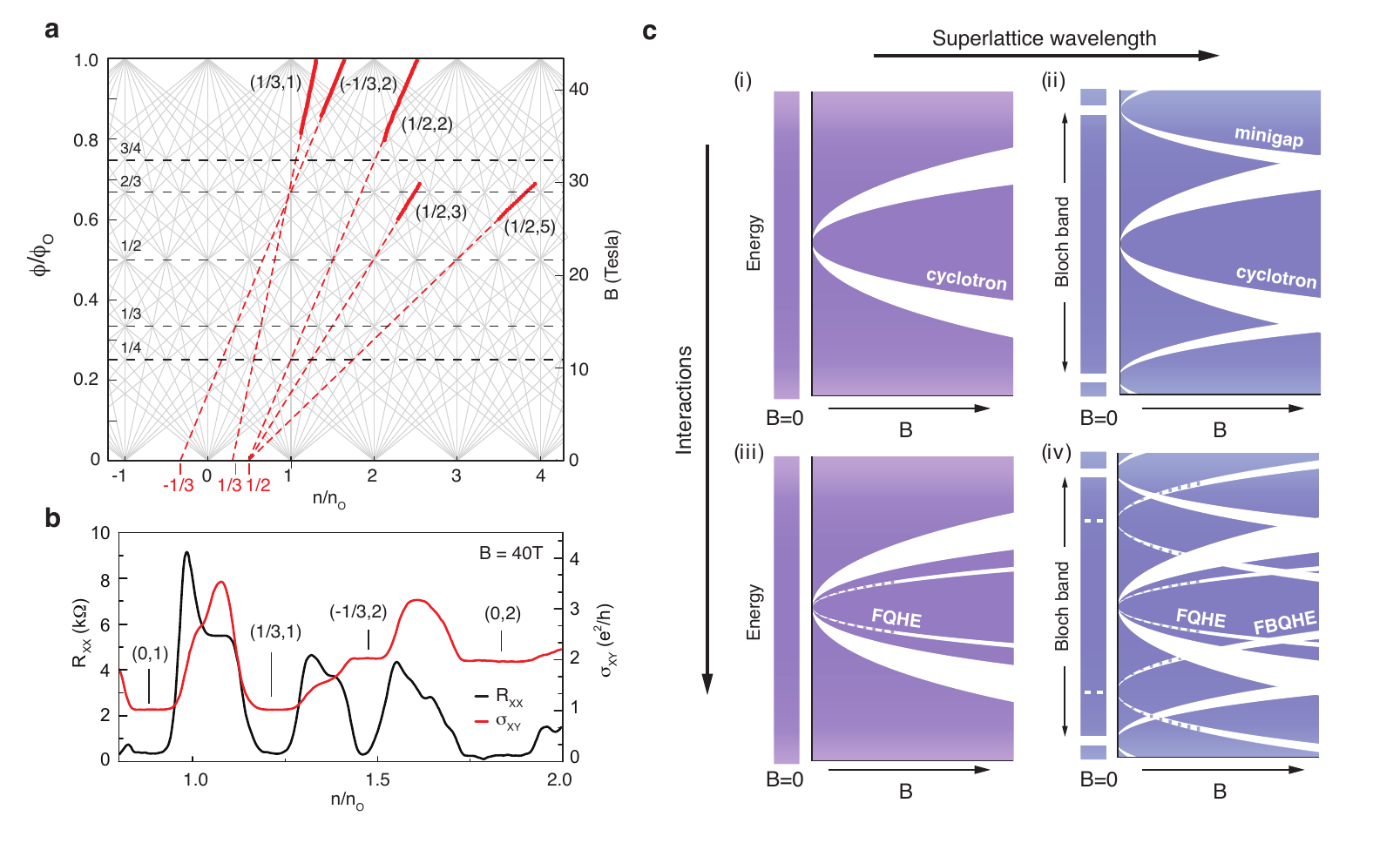}
		\caption{(a) Wannier diagram showing only the anomalous features from Fig 1.  Brackets label the corresponding Bloch band, $s$, and Landau band, $s$, numbers.  Dashed line shows the linear projection to the $n/n_{o}$ axis. (b) Hall conductance and longitudinal resistance versus versus normalized density, measured at fixed magnetic field, $B=40T$ showing transport signatures of select gap states from (a).  The $(s,t)$ numbers determined from the $n/n_{o}$ intercept ($s$) and Hall quantization value ($t$) are labelled in the figure for each QHE plateau. (c) Cartoon summary illustration of the energy bandstructure evolution with magnetic field in the case of no superlattice (left) and with a superlattice (right), and in the limit of no interactions (top) and strong interactions (bottom):  (i) With no superlattice the DOS is continuous at zero magnetic field.  A cyclotron gap develops with finite magnetic field indicated by white against a coloured background.  (ii) In the presence of a large wavelength superlattice, the Bloch band edge is accessible by field effect gating.  Hofstadter minigaps evolve from this band edge, intersecting the conventional Landau levels at large magnetic field. (iii) With no superlattice present, interactions give rise to the fractional quantum Hall effect, appearing also as sub-gaps within the Landau level, but projecting to zero energy. (iv) In the regime of both strong interactions and large wavelength superlattice, we observe a new set of gaps that do not correspond to either the IQHE of single particle gaps, or the conventional many-body FQHE gaps. These are defined by integer valued Hall quantization, but projecting to fractional Bloch band filling indices.}
		\label{fig:Fig3}
	\end{center}
\end{figure*}

Next we discuss the anomalous QHE features associated with the red lines in Fig. 1c. In Fig. 3a, a reduced Wannier plot is shown in which only these anomalous features are replotted (solid red lines), together with a dashed line showing the projection to the $n/n_{o}$ axis. Linear fits to the $R_{XX}$ minimum position versus magnetic field (see SI) indicate that these states follow a trajectory with an integer-valued slope, $t$, but project to non-integer valued intercepts, $s$. Fig. 3b demonstrates unambiguously that these features corresponding to QHE features with well quantized Hall plateaus, and further that the quantization value corresponds to the $t$ value, as expected form the Diophantine equation.  Determining  the fractional $s$ number from the $n/n_{o}$  intercept of the Wannier plot is imprecise  since this depends on calculating the density.  Nonetheless, within experimental uncertainty (see SI), the fractional intercepts appear to cluster around values of 1/3 and 1/2 (Fig. 3a). 

%The anomalous fractional Bloch band QHE (FBQHE) states observed in our study follow trajectories that do not fall within the conventional Wannier diagram, even if allowing for full spin and valley symmetry breaking.  

Fig. 3c shows a cartoon summary of this result.  In the regime of very large magnetic field, in addition to the conventional fractional quantum Hall effect we observe new series of states, outside of the single particle bandstructure as described by the Diophantine equation, and coinciding with a fractional Bloch band index.  The observation of a fractional Bloch band QHE (FBQHE) described by integer $t$ but fractional $s$ numbers may have several possible origins. We note that at $B=30$~T, the Coulomb energy is $\sim80$~meV (($E_{Coulomb}=e^2/\epsilon l_{B}$, where we assume the dielectric constant to be 4).  This is similar in magnitude to the superlattice potential\cite{Moon:2014}, suggesting that interactions play an equivalent role. Recent theoretical consideration of graphene superlattices indeed showed\cite{Ghazaryan:2014} that electron interactions may open a gap  consistent with a fractional $s$ number.  However, the nature of the associated ground state was not identified. Previously, it was predicted that electron interactions may drive a charge density wave (CDW) type modulation of the electron density, commensurate to the superlattice but with a larger period\cite{MacDonald:1985}.  A superstructure with 3 times the  moir\'{e} unit cell area (such as a $\sqrt{3}\times\sqrt{3}$  Kekule distortion) could explain $n/n_{o}$ intercepts of 1/3, whereas a doubling of the superlattice cell could explain 1/2 intercepts. In this regard our observation may resemble the reentrant QHE seen in high mobility GaAs\cite{Eisenstein:2002},  also believed to result from a CDW  phase. Alternatively, one interpretation of the Wannier diagram is to consider the mini-gaps as a sequence of mini Landau fans, residing in a local reduced magnetic field $B'=B-B_{\phi/\phi_{o}=1/m}$, where $\phi/\phi_{o}=1/m$ (or equivalently $1-1/m$ by symmetry) labels regions of high density of mini-gap crossing. Recent bandstructure calculation of moir\'{e}-patterned graphene\cite{ChenUK:2014} indicates  that the mini-fans are not exact replicas but instead can exhibit a local degeneracy with additional Dirac points emerging near $\phi/\phi_o=1/m$.  The FBQHE states may therefore result from an interaction-driven breaking of this degeneracy, similar to quantum Hall ferromagnetism. Finally, we consider that within the mini-fan picture, the FBQHE states resemble the FQHE effect in that they follow trajectories that evolve along fractional filling of the mini-fan LL's, projecting to the $B'=0$ center of the min-fan (see SI). However, both the slope and corresponding Hall conductance plateaux are integer valued.  A complete understanding of our findings will require a theory that accounts for both the observed fractional Bloch band numbers, and simultaneously the associated Hall conductance value. Experimentally, possible ground states could be distinguished by a local probe of the density of states, since for example a CDW phase exhibits broken translation symmetry unlike the Laughlin FQHE state.

\bibliographystyle{naturemag}
\bibliography{Refs}

\bigskip
\section*{Acknowledgments}
We thank A. MacDonald, T. Chakraborty, I. Aeleiner, V. Falko, A.F. Young and B. Hunt for helpful discussions.  A portion of this work was performed at the National High Magnetic Field Laboratory, which is supported by National Science Foundation Cooperative Agreement No. DMR-0654118, the State of Florida and the U.S.Department of Energy. C.R.D. was supported by the US National Science Foundation under grant no. DMR-1463465.

%\section*{Author Contributions}
%L.W., Y.G., B.W., Z.H. and C.R.D. contributed to device fabrication and transport measurement. K.W. and T.T. synthesized the hBN materials. L.W., M.K., J.H. and C.R.D. wrote the manuscript in consultation with all other authors.

%\begin{thebibliography}{10}
%\expandafter\ifx\csname url\endcsname\relax
%  \def\url#1{\texttt{#1}}\fi
%\expandafter\ifx\csname urlprefix\endcsname\relax\def\urlprefix{URL }\fi
%\providecommand{\bibinfo}[2]{#2}
%\providecommand{\eprint}[2][]{\url{#2}}

%\end{thebibliography}

\end{document}